\documentclass{article}
\usepackage{graphicx} 
\usepackage{amsmath}
\usepackage{amsfonts}
\usepackage{xcolor}
\usepackage{authblk}

\title{Resiliency Analysis of LLM generated models for Industrial Automation}

\author[1]{Oluwatosin Ogundare}
\author[2]{Gustavo Quiros Araya}
\author[3]{Ioannis Akrotirianakis}
\author[4]{Ankit Shukla}
\affil{Siemens Technology, Princeton, NJ}

\date{}

\begin{document}

\maketitle

\begin{abstract}
   This paper proposes a study of the resilience and efficiency of automatically generated industrial automation and control systems using Large Language Models (LLMs). The approach involves modeling the system using percolation theory to estimate its resilience and formulating the design problem as an optimization problem subject to constraints. Techniques from stochastic optimization and regret analysis are used to find a near-optimal solution with provable regret bounds. The study aims to provide insights into the effectiveness and reliability of automatically generated systems in industrial automation and control, and to identify potential areas for improvement in their design and implementation.
\end{abstract}
\section{Introduction}
Modelling the interaction of a user with the responses of a Large Language Model(LLM) like ChatGPT to a single question could be conceived as analogous to the multi-armed bandit problem, where the goal is to select the response that maximizes the return. The return in this case is given in terms of a resiliency function and the performance of the user that accepts a particular response from the LLM is measured in terms of the regret, consistent with the prevailing theories in statistical and adversarial bandit problems in online convex optimization. In the classical case, the decision plane engages an exploitation-exploration trade-off with provable regret bounds in the statistical case as shown by Lai \& Robbins \cite{lai1985asymptotically} when strong assumptions is made on the nature of the reward (return) distribution. With no assumptions on the nature of the distribution of the return, the adversarial case also bounds both the weak regret, which in this context is the difference between the cumulative return of an LLM user over a series of questions and the best fixed response acceptance strategy in hindsight as well as a stronger form of the regret that is realized as the difference between the LLM user's cumulative return and the overall best return from the optimal response acceptance strategy \cite{auer2002nonstochastic}. In simple uses of a LLM, like ChatGPT, simply selecting the response that maximizes the Spontaneous Quality (SQ) score might be sufficient for developing an optimal policy but it is insufficient when the reward function is non-trivial \cite{ogundare2023comparative}. The idea of resiliency analysis of automatically generated models is not restricted to derivatives of Large Language Models but also automatically generated scaffolded code and Function Block Diagram (FBD) programs \cite{10097820} \cite{ogundare2023industrial}.\\ \\
Suppose an LLM algorithm uses a finite set of generators
\begin{equation}
    G = \{g_{i}\}
\end{equation}
Such that the automatically selected initial generator is the one that has the highest cumulative empirical rating, $E \in \mathbb{R}$. \\ \\
The empirical rating, $E$ is Bernoulli random variable such that\\
\begin{equation}
   E =\begin{cases}
			1, & \text{User accepts LLM response}\\
            -1, & \text{otherwise}
		 \end{cases}
\end{equation}
An autonomous system, $A$, is designed to accept a singular generator by assigning an empirical rating of $1$ to the generator with the most return, $g^{*}_{i} \in G$, and $-1$ to $\{ G - g^{*}_{i}\}$. The autonomous system, $A$, achieves this by regret minimization. In other words, the generator, $g^{*}_{i}$ produces the response with the smallest associated regret. \\ \\
Generally, the nature of this type of Regret achieved by a first order optimization of the resiliency reward function is proportional to a polynomial in the number of iterations \cite{hazan2007logarithmic}. Such that for every prompt broadly speaking, the resiliency reward function is given by:
\begin{equation}
    \phi(f(g_{i})) \rightarrow \mathbb{Z}^{+}
\end{equation}
\section{Prevailing Theories from Online Optimization}

\subsection{Multi-Armed Bandits (MAB)}
MAB can be thought of as a sequence problem that can be thought of as the discrete analogue of the online optimization problem (described in more detailed in the next subsection). MAB describes a decision process where at each time stage $t$, the decision maker (aka, agent) selects an action $a_t\in A$ where $A=\left\{ 1,2,\dots, |A| \right\}$ is the set of actions available to the agent (in the original MAB context, $A$ is the set of one-arm-bandits the agent can choose from). Once the action is taken, the agent receives a reward $u_t(a_t)$, the time step increases to $t+1$, and the agent can select another action $a_{t+1}\in A$. Based on this simple framework and the way the reward function is defined, we can identify two main types of MABs.

\begin{itemize}
    \item \textit{Stochastic bandits:} its action (or arm) $a\in A$ follows a probability distribution $P_a$ and the reward $u_t(a)$ of action $a$ is a random variable, $u_{t,a}$, drawn from $P_a$. The distribution of the rewards for each action are not known to the agent a priori. The objective is to compute the action (arm) that returns the largest expected reward in as few trials as possible. 

    More specifically, the agent's aim is to select the arm with the highest mean reward, most of the times. If $\mu_a$ is the mean of the reward probability distribution $P_a$ for arm $a\in A$, then the bandit's largest mean reward and the arm that achieves it are respectively defined as 
    \begin{equation}
        \mu^* = \max_{a\in A} \mu_a \quad \mbox{ and } \quad  a^* = \arg \max_{a\in A} \mu_a.
    \end{equation}

    It is frequently assumed that the reward can be drawn from the interval $[0,1]$ and is motivated by the Bernoulli bandit problem where each arm $a\in a$ returns a reward of 1 with probability $p_a$ and 0 with probability $1-p_a$.

    Next we can determine the agent's \textit{mean regret} as the difference between the bandit's best arm and the arm $a_t$ chosen at the time step $t=1,2,\dots,T$. The mean regret is mathematically defined as follows:
    \begin{equation}
        \bar{R}_T = \sum_{t=1}^{T} E[u_t (a^*) - u_t((a_t)] = T\mu^* - \sum_{t=1}^{T} E[u_t((a_t)]
    \end{equation}
    
    An alternative definition of the mean regret is as follows
    \begin{equation}
        \bar{R}_T = E[ \sum_{a\in A} n_{a,T} ] \mu^* - E[ \sum_{a\in A} \mu n_{n,T}]
    \end{equation}
    where $n_{a,T}$ represents the number of times that  $a$ has been selected up to time $T$, the end of our time horizon. 

    \item \textit{Adversarial bandits}:the reward of each action (arm) $a\in A$ is determined by an adversary who decides the rewards to be described by the vector $u_t=(u_t(a))_{a\in A}$. These type of bandits are considered as the most general formulations since they do not consider any probability distributions as in the case of Stochastic bandits.

    In an adversarial bandit problem, at every step the rewards $u_t(a)$ for each arm $a$ are computed by the adversary at the same time the agent decides what action $a_t$ to take. Hence the agent's realized regret (at the end of the time horizon $T$) is defined as
    \begin{equation}
        R_T = \max_{a\in A} \sum_{t=1}^{T}\left[ u_t(a) - u_t(a_t) \right]
    \end{equation}
    which can be interpreted as the cumulative reward difference between the agent's chose arm and the best possible arm over the time horizon we consider.
\end{itemize}

\subsection{Online Convex Optimization}

\section{Optimizing Resiliency by Regret Minimization}

Suppose we have an industrial system composed of subsystems of electro-mechanical components to be actuated by a variety of Siemens PLC apparatus. An intrinsic assumption is that the system is subject to random failures of different types, and we denote by $M$ the number of failures that can occur before the system becomes non-functional. A design assumption is that possibilities for fault tolerance is feasible, at least, in part by a well designed control algorithm. The task of imagining the control surface and designing a control algorithm is assigned to a pre-trained LLM. The performance of a candidate control algorithm is measured by its resilience against potential failures. However, even with central focus on the nature of the algorithms, the notion of resiliency admits more aspects than is covered in this paper. For example, consider the impact of uncertainty in the measurements from components of an industrial system on the decisions made by the control algorithm. Generally speaking, monte carlo methods and method of moments are used to measure uncertainty propagation to improve the decisions made by the control algorithm. However, ironically, strong assumptions are made in practise to make these types of algorithm computationally efficient. For example, using method of moments, the higher order terms of the Taylor series expansion in computing the statistical expectation of a reference independent variable are neglected or assumed to approach a roughly computed empirical limit. What's the impact of these types of practical choices on the liminal resiliency of the system \cite{modarres2006risk}\cite{jasour2021moment}? Regardless, we will focus on the resiliency achievable from the action on an LLM with limited access to dense statistics on the nature and distribution of the model parameters but with a strong understanding of the system, the environment in which it operates and the associated limitations.\\ \\
Let $\phi$ be such that it measures the resilience of the system and is the so-called resiliency reward function, which is given by:
\begin{equation}
\phi(x)= \text{No. of spanning clusters in a 2D lattice graph derived from design x}
\end{equation}
The objective is then to select from a set of designs generated by the LLM, the design that maximizes the resiliency reward function subject to engineering, economic and social constraints, such as the number of modules, the budget, prevailing government policy etc. As stated prior, the regret measures the loss incurred by choosing a sub-optimal design.\\ \\
Let $\phi^*(x)$ be the maximum resilience achievable by any of the LLM generated design. We define the regret of design $x$ as:
\begin{equation}
\text{Regret}(x)=\phi^*(x) - \phi(x).
\end{equation}
The goal is to minimize the maximum regret over all designs, which is given by:
\begin{equation}
\text{Regret}^*(x) = \min_{x} \{\text{Regret}(x)\}.
\end{equation}
Consequently, the regret bound is defined as follows:
\begin{equation}
\text{Regret Bound} = \max_{x} \{\text{Regret}(x) - \text{Regret}^*(x)\}.
\end{equation}
Also we consider the notion of the weak regret, which is the difference between using the best fixed deterministic design strategy and the chosen LLM design.
\section{Deriving a resiliency reward function using percolation theory}

At a very abstract level, we decompose every design $x \in X$ generated by the LLM in response to a reference industrial design question as a set of distinct notions, such that 
\begin{equation}
    x = \{\psi_{i}\},
    \hspace{3mm} i = \{ 1, ..., l\}
\end{equation}
We introduce a lattice generating function, $\psi$, such that
\begin{equation}
    \pi: \{\psi_{i}\} \rightarrow L(\Bar{K}_{m\times n})
\end{equation}
$$ L(\Bar{K}_{m \times n}) \text{ is the edgeless lattice graph of the set of vertices} \{\psi_{i}\} $$
\\
In line with the standard model of percolation theory, every pair of vertices, $\psi_{i}$,$\psi_{j} \in x$ has an Euclidean distance of 1 \cite{Steif2011AMC}. Subsequently, in the geometric rendition of the lattice, edges appear on the lattice with a probability, $p$ estimated from the inferred connectivity of the notions extracted from the generated design. The following ideas follows from the theory and has an intuitive practical implication, 
\begin{itemize}
  \item Whenever $p \geq p_{c}$, where $p_{c}$ is the critical bond percolation probability a spanning cluster exists and the generated system is expected to at least be functional.
  \item The resiliency reward function is now defined in terms of the number of spanning clusters in every generated design $x \in X$
\end{itemize}
Suppose that $\lambda$ is pairwise constant, such that, for any two vertices $\psi_{i}$,$\psi_{j}$
\begin{equation}
    \lambda(\psi_{i}, \psi_{j}):\begin{cases}
			p, & \text{when the edge is open}\\
            1 - p, & \text{otherwise}
		 \end{cases}
\end{equation}
Therefore we compute the percolation probability, $\mathbb{P}_{p}$ as follows:
\begin{equation}
    \mathbb{P}_{p} = \prod_{i, j \in \{1, ..., l \}} \lambda(\psi_{i}, \psi_{j})
\end{equation}
Let $P_{\infty}$ access the probability that an edge belongs to a spanning cluster when it takes the value of 1 otherwise 0. Therefore we choose, $S$ such that
\begin{equation}
    S = \{ \lambda(\psi_{i}, \psi_{j}) \hspace{1mm}|\hspace{1mm} P_{\infty}(\psi_{i}, \psi_{j}) = 1 \}
\end{equation}
$$ i, j = \{ 1, ..., k\} \text{ such that } k \leq l$$ 
The bond percolation model is given as follows
\begin{equation}
    \theta(p) =  \prod_{i \in \{1, ..., k \}} S_{i}
\end{equation}
It worth mentioning that the bond percolation probability is an increasing function of $p$ \cite{Steif2011AMC}. Consequently, we can calculate the critical probability $p_{c}$ as the least upper bound of the set of $p$ for which $\theta(p) = 0$ in $\mathbb{R}^{2}$.

\begin{equation}
    p_{c} = Sup \{ p: \theta(p) = 0 \}
\end{equation}

Practically, there is a proven value for $p_{c}$ in $\mathbb{R}^{2}$ that can be used as a basis for computation.
To find an expression for the average number of spanning clusters, we refer to Fortunato et al\cite{fortunato2004number}. $<k>$ in dimension 2 and it turns out to be proportional to $L^{2 -3y}$ and y is bounded as follows $y \le \frac{2}{3}$ in $\mathbb{R}^{2}$. Similarly, for practical consideration, we conjecture that $y = \frac{1}{2}$ will be suitable for all generated design $x \in X$, such that the average number of spanning clusters approaches the practical limit of $l^{\frac{1}{2}}$.\\

Therefore, we re-state the resiliency reward function formally as
\begin{equation}
    \phi(x) \rightarrow <k>_x
\end{equation}
And the associated regret with design x, as follows
\begin{equation}
    Regret(x) = l^{\frac{1}{2}} - \phi(x)
\end{equation}
Recall, that $l$ is the cardinality of notion in design, x.

\section{CASE STUDY: Analysis of a system of coupled triplex pump, a mixing tank and fast-acting H2O valves designed by ChatGPT}

Prompt: Two chemicals are pumped in by a system of coupled triplex pump, a mixing tank and fast-acting H2O valves are actuated to regulate the concentration of the mixture

\begin{figure}[ht!]
    \centering
    \includegraphics[width=\textwidth]{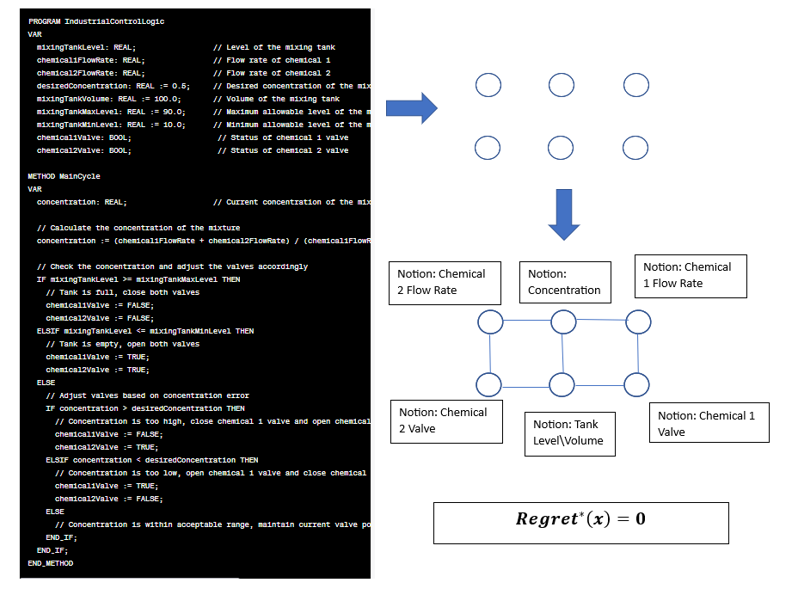}
    \caption{A simple evaluation of a Chemical mixing prompt response using methodology.}
    \label{fig:overview}
\end{figure}

Figure 1 shows how the proposed methodology evaluates a solution obtained from a Large Language Model against its objectives and estimates the theoretical regret against an ideal solution that satisfies all the solution constraints. In this examples case, the automation requirement were completely satisfied within the context of how the PLC actuates the components in the automatic chemical mixing system. The $Regret = 0$ is a practical value in this case because any other structured text or ladder logic code that satisfies the requirement will replicate identical functionality by the PLC. In this sense, there will be no regret incurred in selecting this option over another implementation even if more efficient code in terms of the actual program structure is feasible.

\section{Conclusion \& Further Work}
The field generated by an arbitrary prompt and the idea of the best response has been investigated and a method for formal analysis has been presented in this paper to find a response with the smallest regret. What should follow is an investigation of the relationship between the solution space generated by a prompt perhaps using a finite state machine that shows transitions from the solution with the most regret to the least regret while describing the practical significance of these transitions. Furthermore, it is unclear whether the idea presented in this paper can scale to a cluster of LLMs with a single user interface.
\\
\bibliographystyle{plain} 
\bibliography{refs} 
\end{document}